\documentclass[prb,twocolumn,showpacs,floatfix]{revtex4}
\usepackage{amsmath}
\usepackage{amssymb}
\usepackage{amsfonts}
\usepackage{bm}
\usepackage{graphicx}

\begin{document}
\title{Transitions from small to large Fermi momenta
in a one-dimensional Kondo lattice model}
\author{Eugene Pivovarov}
\author{Qimiao Si}
\affiliation{Department of Physics and Astronomy, Rice University, Houston,
TX 77005}
\date{\today}

\begin{abstract}
We study a one-dimensional system that consists of an electron gas coupled
to a spin-1/2 chain by Kondo interaction away from half-filling.
We show that zero-temperature transitions between phases with
``small'' and ``large'' Fermi momenta can be continuous.
Such a continuous but Fermi-momentum-changing transition arises 
in the presence of spin anisotropy,
from a Luttinger liquid with a small Fermi momentum to
a Kondo-dimer phase with a large Fermi momentum. 
We have also added a frustrating next-nearest-neighbor interaction
in the spin chain to show the possibility
of a similar Fermi-momentum-changing transition,
between the Kondo phase and a spin-Peierls phase,
in the spin isotropic case. This transition, however, 
appears to involve a region in which the two phases coexist.
\end{abstract}


\maketitle

\section{Introduction}

One of the key issues in the theory of heavy fermion 
materials is the size of the Fermi surface.

It is traditional to think that each
spin-$1/2$ localized moment,
through the Kondo effect, contributes a Kondo resonance.
These Kondo resonances, which are fermionic, combine with 
the conduction electrons to form (very heavy) quasiparticles.
In this way, the local moments ultimately become 
a part of the low-energy
electronic fluid.\cite{Hewson} This is reflected in 
the size of the Fermi surface: the Fermi momentum $k_{F}^{\ast}$ is
``large'' in the sense that the Fermi surface encloses a volume
that counts the number of both conduction
electrons and local moments.\cite{Hewson,Oshikawa}
This is in contrast to a ``small'' Fermi momentum $k_F$, which would
correspond to a Fermi surface that has a volume counting only the number
of conduction electrons.
While some questions have been occasionally raised, including
the exhaustion problem noted by Nozi{\`e}res,\cite{Nozieres} this
picture has been thought to apply everywhere in the heavy fermion
phase diagram.\cite{Hewson,Doniach,Auerbach86,Millis87}

Recently, the question about the size of the Fermi surface has 
gained renewed importance in the light of the anomalous behavior
observed in the vicinity of the
antiferromagnetic quantum critical point (QCP).\cite{Stewart}
Particularly noteworthy are the
experiments in 
CeCu$_{\rm 6-x}$Au$_{\rm x}$,
YbRh$_{\rm 2}$Si$_{\rm 2}$,
and related systems,\cite{Schroeder00,Kuchler03,Aronson95}
which observed non-Gaussian behavior. These experiments 
have inspired the development of
locally critical quantum phase transitions\cite{Si01} and
related theoretical pictures.\cite{Coleman01}
In these pictures, the paramagnetic heavy fermion metal phase,
as usual, contains fully developed Kondo resonances, with
a large Fermi momentum $k_F^{\ast}$.
It is argued that the Kondo effect can be destroyed as 
the system approaches a QCP
and goes into an antiferromagnetic metal phase,
by some competing processes: the local moments in such
a lattice setting are also coupled to some fluctuating magnetic 
field, which impedes the Kondo effect.
Once the Kondo effect
is destroyed, the local moments are no longer a part of the 
electron fluid and the Fermi surface becomes small.
Other theoretical works have proposed paramagnetic phases
in which the Kondo effect is destroyed due to the fractionalization
taking place in the localized-spin component.\cite{Senthil,Burdin}
One of the important questions that these theoretical developments
raised is whether quantum transitions between phases with very
different Fermi momenta is necessarily first order or can be 
continuous. Only in the latter case can transitions between such
phases serve as a mechanism for the 
non-Fermi liquid behavior observed
in the heavy fermion metals.

In this paper we address this issue in models in one dimension,
where they are amenable to controlled computation.
We show that continuous transitions from 
small to large Fermi momenta can indeed take place 
in a one dimensional Kondo lattice model away from half-filling.

We now discuss a few additional background issues before 
we go into the details of our analysis.

\subsection{Small and large Fermi momenta}

We prefer to speak about small and large Fermi momenta
as opposed to small and large Fermi volumes. 

The situation in the paramagnetic phase is straightforward
and is standard in the heavy fermion 
literature.\cite{Hewson,Oshikawa}
Whenever the Kondo effect takes place, 
the localized spins are part of the low-energy electron
fluid. The Fermi momentum in this case
is the same as that when each localized spin
is replaced by an electron orbital and, in addition,
with all two-particle 
interactions turned off (i.e., a non-interacting
Anderson lattice model). We call this Fermi momentum
a \emph{large\/} Fermi momentum. On the other hand, in the absence
of the Kondo effect, localized spins --- being charge neutral ---
are not part of the electron fluid. The Fermi momentum is that
of the conduction electrons alone with all interactions turned
off. This is defined as a \emph{small\/} Fermi momentum.
Clearly the Fermi volume (more precisely, Fermi length
in one dimension and Fermi area in two dimensions) is different
in the two cases. A transition between a 
large-Fermi-momentum phase and a small-Fermi-momentum
phase involves a jump in Fermi volumes. 
(This is distinct from other electronic topological
transitions.\cite{Blanter94} For example, in the case
of Lifshitz transitions in metals,
the Fermi volume changes continuously, while the behavior of the
density of states is anomalous.)

The situation is more subtle in an ordered phase that
breaks translational invariance. This includes states 
with either an antiferromagnetic order or a spin-Peierls order.
In the case of even-integer-fold 
breaking of discrete translational invariance
of the Hamiltonian, the Fermi volumes (Fermi area in two dimensions
and Fermi length in one dimension) in the reduced Brillouin zone
would be the same regardless of whether the local moments are a 
part of the electron fluid or not.

However, for commensurate ordering, large Fermi momentum
and small Fermi momentum can still be defined as follows.
The key distinction is again whether local moments are a part
of the electron fluid. We define a large Fermi momentum to be
one that is associated with each localized spin being replaced
by an electron orbital, all two-particle interactions turned
off, but in the presence of an infinitesimal static field that 
characterizes the symmetry breaking. (In a commensurate
antiferromagnetic metal phase, this field would be a static 
staggered magnetic field.) Likewise, we define a small Fermi
momentum to be that associated with conduction electrons
alone, with all two-particle interactions turned off, and
in the presence of the same infinitesimal static field.
In dimensions higher than one, the corresponding large
Fermi surface and small Fermi surface have different
topologies and, hence, represent distinct 
quantum phases.\cite{Blanter94}
In one dimension, on the other hand, the only distinction between
the two is the actual location of the Fermi momentum;
the two phases may not necessarily be distinct if there are
no additional order parameters that differentiate them.

\subsection{The one dimensional Kondo lattice}

The system of interest consists of a one-dimensional electron
gas (1DEG) coupled to a one-dimensional spin-$1/2$ chain 
via Kondo interaction. The spins
in the spin chain $\bm{\tau}_{j}$ interact with each other by
nearest-neighbor (NN) and next-nearest-neighbor (NNN) interactions.
The 1DEG is characterized by Fermi energy $\epsilon_{F}$
and Fermi momentum $k_{F}$.
We take the lattice constants for both the conduction electrons and spin
chains to be equal to $a$. Away from half-filling,
$k_{F}\neq\pi/2a$ so that the 1DEG and the spin chain are incommensurate
with each other.

The nature of Fermi momentum in various phases of 
such a one-dimensional Kondo lattice has been addressed in the past.
That the Fermi momentum in the Kondo phase is indeed $k_{F}^{\ast}$
is supported by a generalized Luttinger's theorem,\cite{Yamanaka97}
derived in a way analogous to that for the Lieb-Schultz-Mattis theorem,
and by the numerical calculations.\cite{Tsunetsugu97,Shibata97}
However, indications for phases with small Fermi momentum have been
shown in more recent 
density matrix renormalization group
calculations\cite{Xavier02} and the effects of spin dimerization
in this context have been recently discussed.\cite{Xavier03}

In absence of Kondo interaction, the 1DEG and 
the spin chain are completely decoupled. In order to study 
the low-energy excitations, one can take the continuum limit
so that 1DEG will be described as a Luttinger liquid.

For large NN interaction, the spin chain will also become a chargeless
Luttinger liquid, with NNN interaction serving as a perturbation. 
When the perturbation becomes sufficiently large, the
system undergoes a phase transition into the spin-Peierls phase.%
\cite{Haldane82}
When the NNN interaction becomes much larger than the NN interaction,
the spin-Peierls ordering persists, but it becomes more convenient to
describe the system in terms of a ``zigzag'' 
model,\cite{White96,Sikkema-PhD}
in which the odd spins and the even spins form two separate chargeless
Luttinger liquids and NN interaction introduces a weak coupling
between them.

The spin-Peierls phase is characterized by staggered bond order parameter,
$\left\langle \bm{\tau}_{j}\cdot\bm{\tau}_{j+1}\right\rangle \sim\left(
-1\right)  ^{j}$. Thus, the spins form dimers,
the translational symmetry is broken, and in certain limits one can even
represent the ground state as a chain of singlets formed by nearest-neighbor
spins. This phase has an exponentially small spin gap. The 1DEG remains a
Luttinger liquid with Fermi momentum $k_{F}$.

Non-vanishing Kondo interaction introduces a number of novel phases in the
system,\cite{Tsunetsugu97,Zachar01b,Zachar96,Novais02}
however, we will be primarily 
interested in limit where the Kondo interaction is weak.
In this case and for an incommensurate lattice, a relevant Kondo coupling
leads to a phase\cite{Zachar01a} in which the electrons and
some of the spins form Kondo singlets $\bm{s}\left(
x_{j}\right)  \cdot\bm{\tau}_{j}$, where $\bm{s}\left(  x\right)  =\frac
{1}{2}\psi^{\dag}\left(  x\right)  \bm{\sigma}\psi\left(  x\right)  $ and
$\psi\left(  x\right)  $, $\psi^{\dag}\left(  x\right)  $ are the
annihilation and the creation operators in the 1DEG. The Kondo singlets are
further dimerized so that the phase is characterized by an order parameter
$\Phi_{\text{K}}$.

Due to the coupling to the charge sector, this order is associated with
gapless charge-density modes $\mathcal{O}$ at $2k_{F}^{\ast}$, 
\begin{equation}
\mathcal{O}\left( x_{j}\right) = \bm{s}\left( x_{j}\right)\cdot\bm{\tau }_{j},
\label{eqn:mode-operator}
\end{equation}
so that the expectation value of this operator varies as $\langle\mathcal{O}\rangle
\sim e^{i2k_{F}^{\ast}x_{j}}$, where $k_{F}^{\ast}=k_{F}+\pi/2a$ 
[a more detailed expression will be given in 
Eqs.~(\ref{eqn:KD-order-parameter},\ref{eqn:charge-mode})
later in the text].
For these modes the charge and spin quantum numbers are both
zero. Note that the charge-density modes are described
by a local operator, however, due to the fact that they embed
the Kondo singlets, the order parameter $\Phi_{\text{K}}$ becomes nonlocal.
As a result, the translational symmetry of the system \emph{is not broken\/}
and the order is hidden. Furthermore, there is an
exponentially small spin gap,\cite{Sikkema-PhD,Sikkema97}
which exists due to the fact that the order parameter is nonlocal and
one cannot distinguish paired spins from unpaired ones.
The existence of the gapless charge-density 
modes at $2k_{F}^{\ast}$ implies
that the system should be described as one with a large Fermi momentum.

We will argue in the following for a new co-existence phase.
In the Kondo dimer phase the spins can further dimerize,
producing a mixed phase in which the Kondo dimer order coexists with
spin-Peierls order. This phase has both a broken translational 
symmetry and a large Fermi momentum.

In addition, we will show that, away from the spin-isotropic case,
the decoupled Luttinger liquid phase has a finite range of stability.
This phase is paramagnetic and has a small Fermi momentum $k_F$.

We now proceed to the quantitative analysis of the
model, which appears to possess all the mentioned phases.

\section{Model}

The Hamiltonian of the system is
\begin{equation}
\mathcal{H}=\mathcal{H}_{\text{1DEG}}+\mathcal{H}_{\text{nn}}+\mathcal{H}%
_{\text{nnn}}+\mathcal{H}_{\text{anis}}+\mathcal{H}_{\text{K}},
\end{equation}
where $\mathcal{H}_{\text{1DEG}}$ is the kinetic term of non-interacting
one-dimensional electron gas (1DEG) with Fermi energy $\epsilon_{F}$,
$\mathcal{H}_{\text{nn}}$ is Heisenberg nearest-neighbor interaction between
the spins in the one-dimensional $1/2$-spin chain,
\begin{equation}
\mathcal{H}_{\text{nn}}=J_{1}\sum\limits_{j}\bm{\tau}_{j}\cdot\bm{\tau
}_{j+1},
\end{equation}
$\mathcal{H}_{\text{nnn}}$ is Heisenberg next-nearest-neighbor interaction in
the spin chain,
\begin{equation}
\mathcal{H}_{\text{nnn}}=J_{2}\sum\limits_{j}\bm{\tau}_{j}\cdot\bm{\tau
}_{j+2},
\end{equation}
$\mathcal{H}_{\text{anis}}$ introduces anisotropy in the spin chain,
whose exact form is not essential provided that the spin chain
by itself ($\mathcal{H}_{\text{nn}}+\mathcal{H}_{\text{nnn}}
+\mathcal{H}_{\text{anis}}$) is gapless. We can, for instance, take 
$\mathcal{H}_{\text{anis}} = - \lambda J_{1}
\bm{\tau}_{j}^{z} \bm{\tau}_{j+1}^{z}$, with $0< \lambda < 1$,
which is an easy-plane ($XY$) anisotropy.
[For a discussion of the role of anisotropy,
see Appendix~\ref{sec:appen-Ising}.]
Finally, $\mathcal{H}_{\text{K}}$ is Kondo interaction between
the 1DEG and the spin chain,
\begin{equation}
\mathcal{H}_{\text{K}}=J_{K}\sum_{j}\bm{s}\left(  x_{j}\right)  \cdot
\bm{\tau}_{j}. \label{eqn:H_K}
\end{equation}
We will focus on the antiferromagnetic case ($J_K > 0$).
As a reminder, here $\bm{\tau}_{j}$ are the spins in the spin chain and
$\bm{s}\left(  x\right)  =\frac{1}{2}\psi^{\dag}\left(  x\right)
\bm{\sigma}\psi\left(  x\right)  $ is the spin density of the electron gas at
$x$. We decompose $\psi\left(  x\right)  $ into right- and left-moving
components, $\psi_{\sigma}\left(  x\right)  =R_{\sigma}\left(  x\right)
e^{ik_{F}x}+L_{\sigma}\left(  x\right)  e^{-ik_{F}x}$,
$\sigma={\uparrow},\downarrow$, so that the spin density becomes
\begin{equation}
\bm{s}\left(  x\right)  =\mathbf{J}_{R}^{s}+\mathbf{J}_{L}^{s}+\mathbf{n}%
^{s},
\label{eqn:s-decompose}
\end{equation}
where $\mathbf{J}_{R}^{s}=\frac{1}{2}R^{\dag}\bm{\sigma}R$, $\mathbf{J}%
_{L}^{s}=\frac{1}{2}L^{\dag}\bm{\sigma}L$, and $\mathbf{n}^{s}=\frac{1}%
{2}R^{\dag}\bm{\sigma}Le^{-i2k_{F}x} +$\ h.~c.

We will study this model in the limit $J_{K},J_{2}\ll J_{1}\ll\epsilon_{F}$.
Then we can take a continuum limit for the spin chain and describe it also
as a one-dimensional free chargeless fermion system with Fermi energy of order
of $J_{1}$. Hence, we can perform a similar decomposition of $\bm{\tau}_{j}$
into right- and left-directed currents $\mathbf{J}_{R}^{\tau}$ and
$\mathbf{J}_{L}^{\tau}$, as well as the staggered spin $\mathbf{n}_{j}^{\tau
}=\left(  -1\right)  ^{j}\mathbf{n}^{\tau}$:
\begin{equation}
\bm{\tau}_{j}=\mathbf{J}_{R}^{\tau}+\mathbf{J}_{L}^{\tau}+\mathbf{n}%
_{j}^{\tau}.
\label{eqn:tau-decompose}
\end{equation}
Likewise, in the limit $J_{1}\ll J_{2}$, the continuum limit of the 
spin chain can be taken using a system
of two mutually noninteracting fermion gases characterized by the Fermi
energies of order of $J_{2}$ (the ``zigzag'' model),
with $J_{1}$ serving as a perturbation.\cite{White96,Sikkema-PhD}

For the single impurity Kondo problem and for the commensurate Kondo lattice
problem, the most relevant part of Kondo interaction is $\mathbf{n}_{j}^{\tau
}\cdot\mathbf{n}^{s}$. However, for the 
incommensurate case --- which is what happens away from half-filling ---
this interaction becomes irrelevant due to the oscillating factors,
and the only component that is not irrelevant is
the forward-scattering one,
\begin{equation}
\mathcal{H}_{\text{K}}^{f}=J_{K}^{f}\sum\left(  \mathbf{J}_{R}^{s}
+\mathbf{J}_{L}^{s}\right)  \left(  \mathbf{J}_{R}^{\tau}
+\mathbf{J}_{L}^{\tau}\right)  ,
\end{equation}
which is marginally \emph{relevant\/} in the $SU\left(  2\right)
$-symmetric case.

Our next step is Abelian bosonization,\cite{Emery79,Gogolin99} in
which we represent the fermionic fields of 1DEG as 
$R_{\sigma}\left(
x\right)  =\left(  F_{\sigma}/\sqrt{2\pi a}\right)  \exp\left\{  -i\sqrt{\pi
}\left[  \theta_{\sigma}-\phi_{\sigma}\right]  \right\}  $ and $L_{\sigma
}\left(  x\right)  =\left(  F_{\sigma}/\sqrt{2\pi a}\right)  \exp\left\{
-i\sqrt{\pi}\left[  \theta_{\sigma}+\phi_{\sigma}\right]  \right\}  $,
$\sigma=\uparrow,\downarrow$, where $\phi_{\sigma}\left(  x\right)  $ are the
bosonic fields, $\theta_{\sigma}\left(  x\right)  =\int_{-\infty}^{x}%
\Pi_{\sigma}\left(  x^{\prime}\right)  dx^{\prime}$ are the fields dual to
$\phi_{\sigma}$, $\Pi_{\sigma}\left(  x\right)  $ are the conjugate fields
satisfying $\left[  \phi_{\sigma}\left(  x\right)  ,\Pi_{\sigma^{\prime}%
}\left(  x^{\prime}\right)  \right]  =-i\delta_{\sigma\sigma^{\prime}}%
\delta\left(  x-x^{\prime}\right)  $, and $F_{\sigma}$ are the Klein factors
satisfying\cite{Kotliar96,Delft98}
$F_{\sigma}^{\dag} F_{\sigma} = 
F_{\sigma} F_{\sigma}^{\dag} =1$ and,
for $\sigma \neq \sigma^{\prime}$, 
$F_{\sigma}^{\dag} F_{\sigma^{\prime}} = -
F_{\sigma^{\prime}} F_{\sigma}^{\dag} $ and
$F_{\sigma} F_{\sigma^{\prime}} = -
F_{\sigma^{\prime}} F_{\sigma}$.
The bosonic fields are further re-expressed in
terms of the spin fields $\phi^{s}\left(  x\right)  =\left(  \phi_{\uparrow
}-\phi_{\downarrow}\right)/\sqrt{2}  $
and charge fields $\phi^{c}\left(  x\right)
=\left(  \phi_{\uparrow}+\phi_{\downarrow}\right)/\sqrt{2} $,
with similar expressions
for their duals $\theta^{s}$ and $\theta^{c}$. The spin chain is bosonized in
a similar manner and we will denote its spin fields as $\phi^{\tau}\left(
x\right)  $.

The charge sector of the model separates from the spin one. Since localized
spins do not carry charge, only 1DEG contributes to this sector, which
is described by a Gaussian model,

\begin{equation}
\mathcal{L}_{c}=\frac{1}{2K_{c}v_{c}}\int dx\left[  \left(  \partial_{\tau
}\phi^{c}\right)  ^{2}+v_{c}^2
\left(  \partial_{x}\phi^{c}\right)  ^{2}\right] .
\end{equation}

Therefore, we will focus on the spin sector. We will study the system in the
general case when the spin chain can be anisotropic. Although we assume that
the microscopic Kondo interaction is isotropic (we will comment on the more
general picture later), it is necessary to keep track of its diagonal
component $J_{K}^{fz}$ separately from the orthogonal component $J_{K}%
^{f\perp}$. As we will show, in the process of the renormalization-group (RG)
flow, the originally isotropic Kondo interaction usually becomes anisotropic
and restores the isotropy only near the fixed point of the Kondo phase.

The resulting Lagrangian becomes
\begin{equation}
\mathcal{L}^{\text{spin}}=\mathcal{L}_{\text{0}}+\mathcal{L}_{\text{SP}%
}+\mathcal{L}_{\text{K}}^{fz}+\mathcal{L}_{\text{K}}^{f\perp} ,
\label{eqn:L_anisotropic}
\end{equation}
where
\begin{subequations}
\label{eqn:L-terms}
\begin{align}
\mathcal{L}_{\text{0}}  &  =\frac{1}{2K_{s}v_{s}}\int dx\left[  \left(
\partial_{\tau}\phi^{s}\right)  ^{2}+v_{s}^{2}\left(  \partial_{x}\phi
^{s}\right)  ^{2}\right] \nonumber \\
& \quad +\frac{1}{2K_{\tau}v_{\tau}}\int dx\left[ \left( \partial_{\tau}%
\phi^{\tau}\right)  ^{2}+v_{\tau}^{2}\left(  \partial_{x}\phi^{\tau}\right)
^{2}\right]  , \label{eqn:L-terms-a} \\
\mathcal{L}_{\text{SP}}  &  =\frac{J_{SP}}{\left(  \pi a\right)  ^{2}}\int
dx\cos\left(  \sqrt{8\pi}\phi^{\tau}\right)  ,\\
\mathcal{L}_{\text{K}}^{fz}  &  =\frac{J_{K}^{fz}}{2\pi}\int dx\,\partial
_{x}\phi^{s}\partial_{x}\phi^{\tau}, \label{eqn:L-terms-c} \\
\mathcal{L}_{\text{K}}^{f\perp}  &  =\frac{2J_{K}^{f\perp}}{\left(  \pi
a\right)  ^{2}}\int dx\cos\left(  \sqrt{2\pi}\phi^{s}\right)  \cos\left(
\sqrt{2\pi}\phi^{\tau}\right) \nonumber\\
& \qquad\times\cos\left[ \sqrt{2\pi}\left( \theta^{s}-\theta^{\tau}\right)
\right]  ,
\end{align}
\end{subequations}
$v_{s}$ and $v_{\tau}$
are the Fermi velocities of the 1DEG and
the spin chain, respectively, $a$ is the lattice constant, and
$J_{SP}$ is the backscattering interaction,
which favors spin-Peierls phase when it is positive.\cite{Haldane82}
For the spin-isotropic case (when $\lambda=0$),
$v_{\tau}=\pi J_{1}/2$ 
$J_{SP}=f\left(  J_{2}-J_{2}^{\ast
}\right)  \simeq c_{1}\left(  J_{2}-J_{2}^{\ast}\right)  $, where $c_{1}%
\simeq1.72\,\pi^{2}$ and $J_{2}^{\ast}\simeq0.24\,J_{1}$.\cite{Eggert96}
Again, this assumes that $J_{2}\ll J_{1}$. If we increase $J_{2}$ to the point
when $J_{K},J_{1}\ll J_{2}$, it will be more appropriate to describe the
system using the ``zigzag'' representation so that
for both spin sub-chains $v_{\tau}\sim J_{2}$ and $J_{SP}\sim J_{1}$.
In addition, finite $\lambda$ modifies the expressions for both
$v_{\tau}$ and $J_{SP}$.

In general, the value of $K_{\tau}$ reflects the fact that the spin chain is
anisotropic
[we will assume that it has 
$XY$ anisotropy and will generalize to Ising
anisotropy in Appendix~\ref{sec:appen-Ising}].
This substantially affects the scaling
dimensions of the terms in Eq.~(\ref{eqn:L_anisotropic}), in particular,
the backscattering 
and transverse Kondo interactions are no longer marginal.

The Kondo part of this Lagrangian contains a term $\cos\sqrt{2\pi}\left(
\phi^{s}-\phi^{\tau}\right)  \cos\sqrt{2\pi}\left(  \theta^{s}-\theta^{\tau
}\right)  $, which is strongly irrelevant. Thus, we can omit it in
$\mathcal{L}_{\text{K}}^{f\perp}$, leading to:
\begin{equation}
\mathcal{L}_{\text{K}}^{f\perp}=\frac{J_{K}^{f\perp}}{\left(  \pi a\right)
^{2}}\int dx\cos\sqrt{2\pi}\left(  \phi^{s}+\phi^{\tau}\right)  \cos\sqrt
{2\pi}\left(  \theta^{s}-\theta^{\tau}\right) .
\label{eqn:Lagr-Kondo-red}
\end{equation}
This expression is quite remarkable. We observe that exchanging the bosonic
fields with their duals $\phi^{s}\leftrightarrow\theta^{s}$, $\phi^{\tau
}\leftrightarrow-\theta^{\tau}$ and simultaneously replacing $K_{s,\tau}$ with
$1/K_{s,\tau}$ leaves the Lagrangian $\mathcal{L}_{\text{0}}+\mathcal{L}%
_{\text{K}}^{f\perp}$ invariant, but induces changes in $\mathcal{L}%
_{\text{K}}^{fz}$ and $\mathcal{L}_{\text{SP}}$. This should manifest itself
not only in the RG equations, but also in the low-energy physics.

Although interaction $J_{K}^{fz}$ alone is marginal and $J_{K}^{f\perp}$ alone
is marginal or irrelevant, these constants effectively modify the scaling
dimensions of each other so that they both can become relevant.

\section{Renormalization-group equations}
\label{sec:RG}

In order to treat the combined effects of the various interaction
terms appearing in the model 
Eqs.~(\ref{eqn:L_anisotropic},\ref{eqn:L-terms}),
we carry out an RG analysis. We will use a 
Coulomb-gas expansion,\cite{Jose77,Kogut79}
in which the interactions associated with the cosine terms
($J_{K}^{f\perp}$ and $J_{SP}$) appear as fugacities,
and those associated with the quadratic terms
($J_{K}^{fz}$, $K_{\tau}$ and $K_s$) specify stiffness
constants. The quadratic parts of the 
Lagrangian Eqs.~(\ref{eqn:L-terms-a},\ref{eqn:L-terms-c})
can be diagonalized.

The general aspects of the derivation of the RG equations 
are standard\cite{Jose77,Kogut79} and will not be detailed here.
What is nontrivial here, however, is the fact that we have two
kinds of fugacities (those associated with $J_{K}^{f\perp}$
and $J_{SP}$), and the way they are coupled with each other
is the key to the determination of the phase diagram. 
Below, we document some of the details regarding this coupling.

The RG equations are found to be

\begin{subequations}
\label{eqn:RG}
\begin{align}
\frac{dy_{\perp}}{dl} &  =\left(  2-\frac{K_{s}}{2}-\frac{K_{\tau}}{2}%
-\frac{1}{2K_{s}}-\frac{1}{2K_{\tau}}\right)  y_{\perp}\nonumber\\
&  \qquad+uK_{s}K_{\tau}y_{\perp}y_{z},\label{eqn:RG-y_K}\\
\frac{dy_{z}}{dl} &  =uK_{\tau}^{2}y_{\perp}^{2},\label{eqn:RG-y_z}\\
\frac{dy_{SP}}{dl} &  =2\left(  1-K_{\tau}\right)  y_{SP},\label{eqn:RG-y_S}\\
\frac{dK_{\tau}}{dl} &  =\left(  1-K_{\tau}^{2}\right)  y_{\perp}^{2}%
-2K_{\tau}^{2}y_{SP}^{2},\label{eqn:RG-K_tau}\\
\frac{dK_{s}}{dl} &  =\left(  1-K_{s}^{2}\right)  y_{\perp}^{2},
\label{eqn:RG-K_s}
\end{align}
\end{subequations}
where $u=\sqrt{2\pi}v_{s}/\left(  v_{s}+v_{\tau}\right)  $ and we have
introduced the dimensionless coupling constants $y_{\perp}=J_{K}^{f\perp
}/\left(  2\pi\right)  ^{3/2}v_{s}$, $y_{z}=J_{K}^{fz}/\left(  2\pi\right)
^{3/2}v_{s}$, and $y_{SP}=J_{SP}/4\pi^{3/2}v_{\tau}$. Note that the
coefficients of the $y_{\perp}^{2}$ and $y_{SP}^{2}$ terms in
Eqs.~(\ref{eqn:RG-K_tau},\ref{eqn:RG-K_s}) actually depend on the method of
regularization, however, those of the $y_{\perp}^{2}$ term must preserve the
invariance of the equations (for $y_{SP}=y_z=0$) with respect to the change of
variables $K_{s}\rightarrow1/K_{s}$ and $K_{\tau}\rightarrow1/K_{\tau} $ due
to the symmetry of Eq.~(\ref{eqn:Lagr-Kondo-red}) discussed above. At the
spin-chain isotropic point with $y_{SP}=0$, the flow is along $K_{s}=K_{\tau
}=1$. Far away from this point the Kondo interaction is irrelevant and its
flow is dominated by the first term in Eq.~(\ref{eqn:RG-y_K}).

There is no relevant correction due to $y_{z}$ or $y_{\perp}$\ on the
right-hand side of Eq.~(\ref{eqn:RG-y_S}). In the third order of perturbation
we have found the following term there:
\begin{equation}
-2\pi\frac{v_{s}^{2}\left(  v_{s}+2v_{\tau}\right)  K_{s}K_{\tau}^{2}}%
{v_{\tau}\left(  v_{s}+v_{\tau}\right)  ^{2}}y_{z}^{2}y_{SP}.
\end{equation}
A similar expression has been obtained for a different spin-Peierls order
as well.\cite{Sachdev88}
The fact that this correction makes 
the backscattering interaction more irrelevant
means that there is a competition between the 
Kondo-dimer and spin-Peierls phases
and not a mutual enhancement. If there was enhancement
(as in the case of melting in two dimensions), the two primary phases
would be one with both kinds of order being present and one with no order.
In our case, the mixed phase can only be intermediate, while the pure Kondo
and spin-Peierls phases play major roles. Remarkably, in the second order
of perturbation even the direct competition vanishes.

In absence of either the backscattering $J_{SP}$ or transverse Kondo
$J_{K}^{f\perp}$ interactions, the remaining $z$-component of Kondo interaction
$J_{K}^{fz}$ is strictly marginal. However, the theory is still
Gaussian\cite{Zachar01a} and, by using a linear transformation of fields
$\phi^{s}$ and $\phi^{\tau}$, we have obtained
(see Appendix~\ref{sec:appen-small-kF})
an exact solution describing two decoupled liquids
characterized by new spin wave velocities $v^{\prime}_{s}$, $v^{\prime}_{\tau}$,
with corrections of order of $\left( J_{K}^{fz}\right) ^2$ for
$J_{K}^{fz}/2\pi\ll v_{s},v_{\tau}$. Thus,
we still have a Luttinger-liquid state with four 
``Fermi points'' in the spin sector,
two at $\pm k_{F}$ and two at $\pm\pi /2a$, which means that small $J_{K}^{fz}$
does not change the behavior of the system qualitatively.
In each of the decoupled liquids the particles are partially in the spin sector
of the original electron liquid and partially in the original localized-spin
liquid. Increasing $J_{K}^{fz}$ enhances this mixing until one
encounters a singularity, which could be a signature of
a transition into the ``Toulouse point phase''%
\footnote{As we increase $J_{K}$ beyond 
$\epsilon_{F}$ (keeping $J_{1},J_{2}\ll\epsilon_{F}$), we will reach 
a ``Toulouse point'' phase,\cite{Zachar96}
which still has a spin gap and
which is characterized not only by the Kondo dimer 
order parameter but also with ordinary charge-density wave
(with a gapless mode at $2k_{F}^{\ast}$).}
with Fermi momentum $k_{F}^{\ast}$ or the development of quasi-long-range
spin-density-wave order.\cite{Sachdev88}

There is a process that is not being taken into account in Eqs.~(\ref{eqn:RG}).
During the RG flow, a term $\partial_{x}\theta^{s}\partial_{x}\theta^{\tau}$
is generated. This term has a physical meaning of an interaction
between the $z$-components of physical spin currents of the
electrons and the localized moments. Its coupling constant (let us denote it
$\bar{y}_{z}$) renormalizes as in Eq.~(\ref{eqn:RG-y_z}), with the replacement
of $K_{\tau}^{2}$ with $K_{\tau}^{-2}$, and it makes a contribution to
$\beta\left(  y_{\perp}\right) $ similar to one that has already been made by
$y_{z}$. As long as $y_{\perp}$, $y_{z}\ll1$, this correction only negligibly
affects the flow of $y_{\perp}$. However, in the Kondo phase all three
coupling constants become approximately equal to each other and renormalize to
infinity. Taking $\bar{y}_{z}$ into account rigorously might affect certain
numerical factors in the critical exponents, but it should not change the
critical behavior qualitatively, therefore it has been neglected in
Eqs.~(\ref{eqn:RG}).

Let us now focus on the equations Eqs.~(\ref{eqn:RG-y_K},\ref{eqn:RG-y_z}%
,\ref{eqn:RG-K_tau}) involving only Kondo interaction in presence of
anisotropic spin chain (which also corresponds to the case when $y_{SP}=0$ and
$K_{s}=1$). We aim at determining the separatrix 
which divides the region where the Kondo coupling ($y_{\perp}$) is irrelevant 
from the region where it is relevant. For small values of the bare Kondo
coupling, we will see that the anisotropy $x=K_{\tau}-1$ is small
on the separatrix. For small $x$, the RG equations simplify to
the following:
\begin{subequations}
\label{eqn:RG-reduced}
\begin{align}
\frac{dy_{\perp}}{dl}  &  =-\frac{1}{2}x^{2}y_{\perp}+uy_{\perp}y_{z},\\
\frac{dy_{z}}{dl}  &  =uy_{\perp}^{2},\\
\frac{dx}{dl}  &  =-2xy_{\perp}^{2}.
\end{align}
\end{subequations}
These equations have the following first integrals:
\begin{subequations}
\label{eqn:1st-int}
\begin{align}
y_{z}^{2}-y_{\perp}^{2}+\frac{x^{2}}{4}  &  =C_{1},\\
xe^{2y_{z}/u}  &  =C_{2}.
\end{align}
\end{subequations}
The physical meaning of the constant $C_{1}$ can be derived from the initial
condition that the Kondo interaction is isotropic at the beginning of the
flow, when $x=x_{0}$ and $y_{\perp}=y_{z}=y_{0}$, thus, $C_{1}=x_{0}^{2}/4$.
While $C_{2}$ can be understood as the value of $x=\tilde{x}$ such that
$y_{z}(\tilde{x})=0$, it is only a formal definition, since $y_{z}$ never
vanishes or changes sign for the mentioned initial conditions.
Therefore,
$C_{2}\equiv\tilde{x}$ actually determines the value of $y_{0}$.

\begin{figure}[tbh]
\begin{center}
\includegraphics[width=3.25in]
{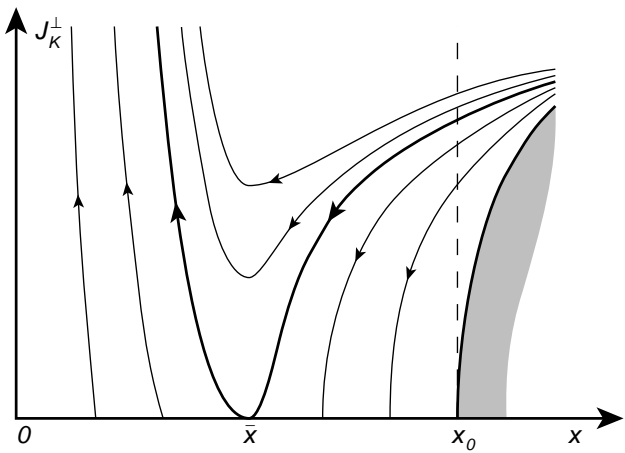}
\end{center}
\caption{The RG flow of $J_{K}^{f\perp}$ \emph{vs.\/} $x=K_{\tau}-1$ for
$J_{SP}=0$. At $x_{0}$ the Kondo interaction is isotropic, which is the
initial condition.}
\label{fig:flow-K}
\end{figure}

The equations Eq.~(\ref{eqn:1st-int}) can be written in a more convenient
form,
\begin{subequations}
\begin{align}
y_{\perp}  &  =\pm\left(  y_{z}^{2}+\frac{x^{2}}{4}-\frac{x_{0}^{2}}%
{4}\right)  ^{1/2},\\
y_{z}  &  =\frac{u}{2}\ln\frac{\tilde{x}}{x},
\label{eqn:yz}
\end{align}
\end{subequations}
which represents the lines of the flow of the RG charges $y_{\perp}$ and
$y_{z}$ (Fig.~\ref{fig:flow-K}). As we see, $dy_{\perp}/dl$ (and consequently,
$dy_{\perp}/dx$) vanishes at $x=\bar{x}$, where
\begin{equation}
\bar{x}=\frac{u}{\sqrt{2}}\left(  \sqrt{1+\frac{4x_{0}^{2}}{u^{2}}}-1\right)
^{1/2}.
\end{equation}
For $x_{0}\ll u$, one can approximate $\bar{x}\simeq x_{0}-x_{0}^{3}/2u^{2}$.
There is also a flow line that separates the lines ending at $y_{\perp}=0$
from the lines flowing towards $y_{\perp}=\infty$. This separatrix is
determined by
\begin{equation}
\tilde{x}_{\text{sep}}=\bar{x}\exp\left(  \frac{1}{u}\sqrt{x_{0}^{2}-\bar
{x}^{2}}\right)  ,
\label{eqn:x-tilda-sep}
\end{equation}
which is $\tilde{x}_{\text{sep}}\simeq x_{0}+x_{0}^{3}/2u^{2}$ for $x_{0}\ll
u$.

It follows that in general, even if initially $y_{\perp}=y_{z}=y_{0}$, the
Kondo interaction becomes anisotropic as a result of the RG flow such that
$y_{\perp}<y_{z}$. In particular, at the line of Luttinger-liquid fixed points
$\bar{x}<x<x_{0}$, $y_{\perp}=0$, the diagonal component of Kondo interaction
$y_{z}$ is finite. However, if the lines flow towards the fixed point of the Kondo
phase $x=0$, $y_{\perp}=\infty$, the isotropy is eventually restored,
$y_{z}\rightarrow y_{\perp}$.

A nontrivial consequence of the presented calculation is that the flow
along the separatrix ends at a point where not only Kondo interaction
is anisotropic, but also spin-chain $XY$ anisotropy is characterized by
a nonzero value $\bar{x}$. This value is nonuniversal and is related
to the strength of the $z$-component of Kondo interaction at the point.
These features are unique to our problem and have to do with the
fact that both $y_z$ and $K_{\tau}$ affect the conditions for the 
development of the Kondo phase. However, since this transition
involves the change of the behavior of one fugacity,
we expect that the universal properties will be of
the Kosterlitz-Thouless type. Indeed, by introducing
$\epsilon=\frac{1}{2}x^{2}-uy_{z}$, the RG 
equations~(\ref{eqn:RG-reduced})
can be expressed in the standard Kosterlitz-Thouless form.


\section{Phases and Fermi momenta}

First, we consider the quantum phases as characterized by the fixed points
of the RG flow.

\subsection{Decoupled Luttinger liquid with small Fermi momentum}

\begin{figure}[tbh!]
\begin{center}
\includegraphics[width=3.25in]
{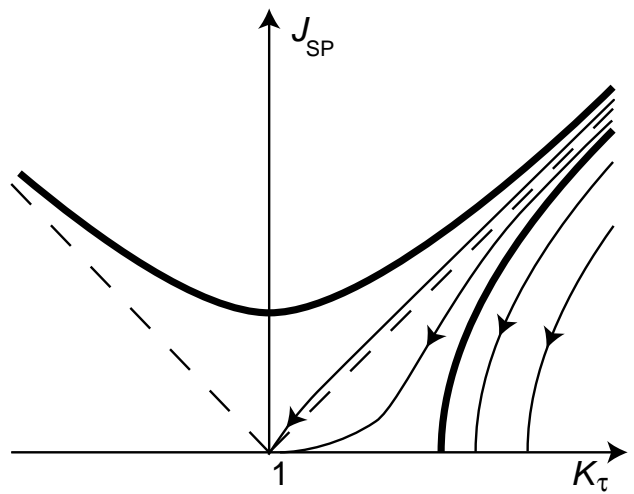}
\end{center}
\caption{The RG flow of $J_{SP}$ for initially small $J_{K}$. The simultaneous
flow along $J_{K}$ is not shown.}
\label{fig:flow-SP}
\end{figure}

For spin-isotropic systems,
the decoupled Luttinger liquid phase is unstable.
The situation is drastically different when we go to the anisotropic case.
For simplicity, we consider $K_s=1$ corresponding to the case in which
the 1D electron gas is non-interacting.
If initially $y_{K}\ll y_{SP}$ and $K_{\tau}\gg1$, the 
backscattering 
interaction $y_{SP}$ rapidly renormalizes
to zero (Fig.~\ref{fig:flow-SP}). After that we only need to consider
the flow of the $y_K$. 
For $K_{\tau} > 1$, there is a finite range of the Kondo interactions
(both $J_{K}^{fz}$ and $J_{K}^{f\perp}$) over which
$y_K$ renormalizes to zero.

The impedance to the Kondo effect here comes from
the formation of a Luttinger liquid among the localized spins.
The latter changes the scaling dimension of the Kondo interaction
and, over an appropriate region of the interaction parameter 
space, renders the Kondo coupling irrelevant in the RG sense.
While the details of this mechanism differ from those of
either Ref.~\onlinecite{Si01} (fluctuating magnetic field) or
Refs.~\onlinecite{Senthil} (gapped spin liquid), the effect
is similar: they lead to the destruction of the Kondo effect.

The only subtlety in the calculation of correlation functions
in this phase is the $J_K^{fz}$-coupling, which is marginal.
It mixes the spin degrees of freedom from the spin chain and from 1DEG.
This mixing can be straightforwardly treated by introducing
a new basis, leading to two decoupled spin branches with renormalized
velocities. The details are given in Appendix~\ref{sec:appen-small-kF}.
The spin excitations are gapless at both $2k_F$ and 
$\pi/a$, which is due to the fact that $J_K^{fz}$ does not induce
a magnetic order and, consequently, does not create new spin
excitations. 

The charge sector is completely independent of the
$J_K^{fz}$-coupling,
therefore, the number of particles in this sector does not change
at all. Consequently, the Fermi momentum is not affected and
the charge excitations are gapless only at $2k_F$. In addition,
the single-electron excitations are gapless only at $k_F$:
while $J_K^{fz}$ changes the shape of the spectral function,
it does not induce gapless single-electron excitations at other
wavevectors. In this sense, the Fermi momentum is small and is 
equal to $k_F$.

We should stress that there is no violation of the 
``generalized Luttinger theorem'' as specified in 
Ref.~\onlinecite{Yamanaka97}. This theorem states that
the twist operation ($U$) will introduce a gapless
state, $U|{\text{gs}}\rangle$, whose momentum differs from that
of the ground state $|{\text{gs}}\rangle$ by $2k_F+\pi/a$. 
In the decoupled Luttinger liquid, a part of the momentum
($2k_F$) measures the change to the 1DEG and
the other part ($\pi/a$) measures the change to the spin chain.

\subsection{The Kondo phase with large Fermi momentum}

Another RG fixed point is given by $J_{K}^{f\perp}=\infty$.
As discussed earlier,\cite{Zachar01a} this 
corresponds to the spin-gap phase with $\sqrt{2\pi}\left(  \phi^{s}+\phi
^{\tau}\right)  =\left(  2n_{1}+1\right)  \pi$, $\sqrt{2\pi}\left(  \theta
^{s}-\theta^{\tau}\right)  =2\pi n_{2}$, where $n_{1,2}$ are integers, or with
$\sqrt{2\pi}\left(  \phi^{s}+\phi^{\tau}\right)  =2\pi n_{1}$, $\sqrt{2\pi
}\left(  \theta^{s}-\theta^{\tau}\right)  =\left(  2n_{2}+1\right)  \pi$. This
phase is precisely the state in which Kondo singlets $\mathbf{n}^{s}%
\cdot\mathbf{n}_{j}^{\tau}$ form an order parameter
\begin{equation}
\Phi_{\text{K}}=\left\langle \cos\sqrt{2\pi}\left(  \theta^{s}-\theta^{\tau
}\right)  -\cos\sqrt{2\pi}\left(  \phi^{s}+\phi^{\tau}\right)  \right\rangle .
\label{eqn:KD-order-parameter}
\end{equation}
This order parameter is associated with the gapless charge-density mode
Eq.~(\ref{eqn:mode-operator}) at $2k_{F}^{\ast}$,%
\footnote{Another
quantity that can be dimerized is $\bm{\Delta}^{s}\cdot\mathbf{n}_{j}^{\tau}$,
where $\bm{\Delta}^{s}=iL^{s}\left(  \sigma_{2}\bm{\sigma}\right)  R^{s}$ is
triplet superconductivity pairing; the associated gapless modes $\Phi
_{\text{KD}}\exp\left[  i\theta_{c}\left(  x\right)  +i\pi x/a\right]  $ have
momentum $\pi/a$.}
\begin{equation}
\mathcal{O}\left( x\right) \sim \Phi_{\text{K}}e^{
i\sqrt{2\pi}\phi^{c}\left( x\right)+i2k_{F}^{\ast}x} .
\label{eqn:charge-mode}
\end{equation}

In the left-hand side of Eq.~(\ref{eqn:charge-mode}),
one can replace $\mathcal{O}\left( x_{j}\right) $ with
$-\mathcal{O}\left( x_{j}\right) \mathcal{O}\left( x_{j+1}\right)$,
which follows from $J^{s+}n^{s-}\sim n^{sz}$ and similar relations.
Physically this new formula means that the spin sector of the Kondo
phase contains the dimers of the Kondo singlets. However,
the order parameter $\Phi_{\text{K}}$ also includes a contribution
from the charge sector $\exp\left( -i\sqrt{2\pi }\phi ^{c}\right) $.

Only relative values of the phase bosonic fields attain
nonvanishing expectation values, while the fields themselves remain
fluctuating. This picture does not depend on whether the original Kondo
interaction was anisotropic or not, thus, \emph{Kondo dimer phase restores the
isotropy of the Kondo interaction}.

\subsection{The spin-Peierls phase with a small Fermi momentum}
\label{subsection:spinpeierls}

Yet another RG fixed point is given by 
$y_{SP}=+\infty$ and $y_K=0$,
corresponding to the spin-Peierls
phase with $\sqrt{2\pi}\phi^{\tau}=\pi n$, where $n$ is an integer. 
Here, the Kondo effect is destroyed due to the spin gap induced by
the spin-Peierls ordering in the spin chain.
As a result, the Fermi momentum is $k_F$.

\subsection{A coexisting Kondo and spin-Peierls phase}
\label{subsection:coexisting}

An interesting possibility is described by the RG fixed point
with $y_K = y_{SP} = + \infty$. This is the regime in which
the system is
simultaneously in the Kondo-dimer and spin-Peierls phases.
Then all fields $\phi^{s,\tau}$, $\theta^{s,\tau}$ attain
nonvanishing expectation values
and the spectrum of spin excitations becomes gapped. There is
still a gapless charge-density mode at $2k_{F}^{\ast}$ in this state, though.

Because of the broken translational symmetry, the coexisting phase
and the purely spin-Peierls phase have the same 
Fermi volume (length). They are, however, distinct phases due to the 
absence (presence) of the Kondo-dimer order parameter and the spin gap
in the spin-Peierls (coexistence) phase.

If we extend our study towards Ising anisotropy of the spin chain
(see Appendix~\ref{sec:appen-Ising}), we will have
to add Ising phase to the total picture as well. However, this phase
is of little interest in the context of our problem.

\section{Quantum phase transitions}

We now consider in some detail the transitions between the phases.
For simplicity, we will again set $K_{s}=1$. 

\subsection{Transition from the decoupled Luttinger liquid to
the Kondo phase}

First, we will study the
transition from the Luttinger-liquid phase to
the Kondo-dimer one. For $y_{SP}<x$,
$x=K_{\tau}-1$, the backscattering 
interaction $y_{SP}$ always renormalizes to
zero. There is, however, a separatrix between the regions where $y_{\perp}$ is
relevant and where it is irrelevant (Fig.~\ref{fig:flow-K}). By substituting
Eq.~(\ref{eqn:x-tilda-sep}) into Eq.~(\ref{eqn:yz}), we find\ that the
equation of the separatrix is
\begin{equation}
y_{K}^{\text{sep}}=\frac{u}{2}\ln\left(  \frac{\bar{x}}{2x}\right)  +\frac
{1}{2}\sqrt{x^{\ast2}-\bar{x}^{2}},
\label{eqn:separatrix}
\end{equation}
where $x^{\ast2}=x_{0}^{2}-y_{SP}^{2}$. Thus, the phase boundary corresponding
to this transition is determined by the initial conditions for isotropic Kondo
interaction $y_{\perp}=y_{z}=y_{K}$ of the form $y_{K}=y_{K}^{\text{sep}%
}\left(  x_{0}\right)  $. An approximate equation of the phase boundary is
$y_{K}\simeq x^{\ast2}/4u$ for $x^{\ast}\ll u$, in agreement with the result
obtained at the end of Sec.~\ref{sec:RG}.

When Kondo interaction is stronger than $y_{K}^{\text{sep}}$, the flow of the
constants $y_{\perp}$ and $y_{z}$ eventually becomes relevant and $x$
approaches zero. Suppose that initially $y_{K}=y_{K}^{\text{sep}}\left(
x_{0}\right)  +t$, where $t\ll x^{\ast}$. Then the flow of $x$ is determined
by the following differential equation:
\begin{multline}
\frac{dx}{dl}=-\frac{x}{2}\left[  u^{2}\ln\left(  \frac{\tilde{x}_{\text{sep}%
}}{x}\right)  ^{2}+x^{2}-x^{\ast2}\right. \\
\left.  +4ut\ln\left(  \frac{\tilde{x}_{\text{sep}}}{x}\right)  \right] .
\end{multline}
\newline By expanding it about $x=\bar{x}$ and assuming that $x^{\ast}\ll u$,
we derive
\begin{equation}
\frac{dx}{dl}=-\frac{2t}{u}x^{\ast3}-\frac{u^{2}}{x^{\ast2}}\left(  x-\bar
{x}\right)  ^{2}.
\end{equation}
The magnitude of the spin gap is determined by the value of the parameter $l$
when $y_{K}$ becomes of order of unity, $\Delta\sim\exp\left(  -l^{\ast
}\right)  $. Thus, we find that in the Kondo-dimer phase near the transition
the spin gap is exponentially small,
\begin{equation}
\Delta_{K}\sim\exp\left[  -\frac{\pi}{x^{\ast}\sqrt{2u\left(  y_{K}%
-y_{K}^{\text{sep}}\right)  }}\right]  .
\end{equation}

Therefore, the transition from Luttinger-liquid phase to the Kondo-dimer phase
is \emph{continuous\/} and, indeed, belongs to the
Kosterlitz-Thouless type.
However, the exponent is not universal due to the dependence on $x_{0}$. The
formula above is correct only for $t=y_{K}-y_{K}^{\text{sep}}\ll x^{\ast}$ and
it becomes invalid as $x^{\ast}\rightarrow0$, where it is $\Delta_{K}\sim
\exp\left(  -1/ut\right)  $.

Now consider the behavior of the correlation functions at the transition. We
observe that the correlation functions $\left\langle \mathbf{s}%
(x,\tau)\cdot\mathbf{s}(0)\right\rangle $ and $\left\langle \bm{\tau}(x,\tau
)\cdot\bm{\tau}(0)\right\rangle $ are nonuniversal. Indeed, the uniform part of
the former decays as $r^{-\gamma_{s}}$ and the uniform part of the latter is
$r^{-\gamma_{\tau}}$, where $r=x\pm v\tau$ and
\begin{subequations}
\begin{align}
\gamma_{s}  & =2+\pi\frac{v_{s}\left(  v_{\tau}+2v_{s}\right)  }{\left(
v_{\tau}+v_{s}\right)  ^{2}}K_{\tau}\bar{y}_{z}^{2},\\
\gamma_{\tau}  & =K_{\tau}+\frac{1}{K_{\tau}}+\pi\frac{v_{s}^{2}\left(
v_{s}+2v_{\tau}\right)  K_{\tau}^{2}}{v_{\tau}\left(  v_{s}+v_{\tau}\right)
^{2}}\bar{y}_{z}^{2}.
\end{align}
\end{subequations}
Here $\bar{y}_{z}=y_{z}\left(  \bar{x}\right)  $ on the separatrix, $\bar
{y}_{z}\simeq x_{0}^{2}/2u$. The staggered part of the localized-spin
susceptibility decays as $r^{-1/K_{\tau}}$, as it involves only $\theta_{\tau
}$ and no $\phi_{\tau}$.
The corresponding local spin susceptibility,
$\chi(\omega; x=0)$, is divergent.
(The divergence at the QCP becomes logarithmic in the spin-isotropic limit,
which is consistent with an extrapolation towards the QCP of the result
obtained inside the Kondo-dimer phase of the spin-isotropic model using
the form-factor technique.\cite{Tsvelik})

In order to construct a correlation function that is characterized by a
universal critical exponent at the transition, we will introduce the following
operator, defined in terms of right- and left-moving fields:
\begin{equation}
P=\sum\limits_{\sigma}
R_{\sigma}^{\tau\dag}L_{\sigma}^{s}+L_{\sigma}^{\tau\dag}R_{\sigma}%
^{s}+\text{h. c.}
\end{equation}
The fields $R_{\sigma}^{s}$ and $L_{\sigma}^{s}$\ are the spin pieces of the
right- and left-moving fermions in the 1DEG, introduced earlier, so that
$R_{\sigma}^{s}$, $L_{\sigma}^{s}\sim\exp\left[  \sigma i\sqrt{\pi/2}\left(
\pm\phi^{s}+\theta^{s}\right)  \right]  $.
This operator mixes the spinons of the spin chain and those
of the conduction electron gas.
While the spin does not flip in the process of such transformation, the
momentum is not conserved and changes by $k_{F}^{\ast}$. Judging by the
representation of $P$ in the bosonic fields, one could formally identify it
with the square root of the Kondo singlet, $P\sim\left(  \mathbf{s}_{\perp
}\cdot\bm{\tau}_{\perp}\right)  ^{1/2}$. The susceptibility associated with $P
$, $\chi_{P}\left(  x,\tau\right)  =\left\langle P\left(  x,\tau\right)
P\left(  0\right)  \right\rangle $, decays as a power law $r^{-1-\epsilon
/2}$, therefore, the corresponding local susceptibility at the transition
(where $\epsilon=0$) has
universal behavior
\begin{equation}
\chi_{P}\left(  \omega;x=0\right)  \sim\ln\left(  \frac{i\Lambda}{\omega
}\right)  ,\quad\omega\rightarrow0,
\end{equation}
where $\Lambda$ is a cutoff.

\subsection{Transition between the Kondo, spin-Peierls,
and coexistence phases}

If we start at the separatrix Eq.~(\ref{eqn:separatrix}) and begin
to decrease $K_{\tau}$, we will get deeper and deeper inside the Kondo
phase. By iterating the RG equations, we found that there are two
additional regions of the phase diagram.

For $K_{\tau}$ sufficiently smaller than the separatrix value,
we find that $y_{\perp}$ renormalizes to zero while
$y_{SP}$ renormalizes to larger and larger values. This yields
the pure spin-Peierls phase discussed in 
Sec.~\ref{subsection:spinpeierls}.

Between the Kondo phase and the spin-Peierls phase, we find a finite
region of parameters over which both $y_{\perp}$ and $y_{SP}$ renormalize
to larger values and reach order unity simultaneously. 
We interpret this as meaning that we enter a state in which the Kondo
dimers \emph{coexist\/} with spin-Peierls phase. 

The exact
locations of the corresponding phase boundaries are impossible to determine
from the RG equations, since the latter become invalid when $y_{\perp}\sim1$
or $y_{SP}\sim1$. However, we can determine an approximate location of the
coexistence phase from the condition that both $y_{\perp}$ and $y_{SP}$
reach $O(1)$ at certain finite value of $l$. This line has the
following shape for finite $y_{K}$:
\begin{align}
y_{SP}  &  =x+\left(  \frac{\pi}{2}uy_{K}\right)  ^{2},\quad x>0,\nonumber\\
&  \sim\exp\left(  \frac{2}{uy_{K}}\left\vert x\right\vert \right)  ,\quad
x_{c}<x<0, \label{eqn:dashed-line} \\
&  \sim\exp\left\{  -\left[  \frac{2\pi^{2}}{u\left(  y_{K}-\frac{x^{2}}%
{4u}\right)  }\right]  ^{1/2}\right\}  ,\quad x\rightarrow x_{c},\nonumber
\end{align}
where $x_{c}=-2\sqrt{uy_{K}}$. The dependence of the spin-gap in the
spin-Peierls phase near $x_{c}$ is $\Delta_{SP}\sim y_{SP}^{1/4\sqrt{uy_{K}}}
$.

The same line for finite $y_{SP}$ ends at the point $x=y_{SP}$, at which all
three phase boundaries merge and which separates spin-Peierls phase from
Luttinger liquid when $y_{K}$ vanishes.

\subsection{Phase diagram}

\begin{figure}[tbh]
\begin{center}
\includegraphics[width=3.25in]
{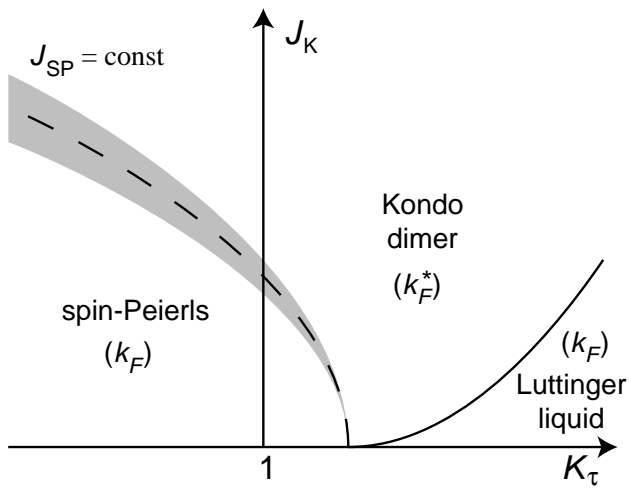}
\end{center}
\caption{Phase diagram for fixed finite $J_{SP}$. 
The solid line shows the continuous transition and
the shaded area shows the possible location
of the coexistence region. The dashed line has been defined in Eq.~(\ref{eqn:dashed-line}), but the exact shape of the area
is unknown yet.
The lines merge at a single point. The brackets label the Fermi
momentum of each phase.}
\label{fig:phase-K}
\end{figure}

\begin{figure}[tbh!]
\begin{center}
\includegraphics[width=3.25in]
{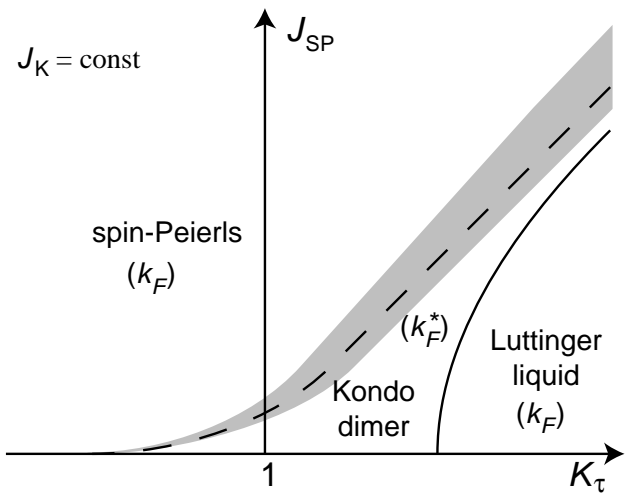}
\end{center}
\caption{Phase diagram for fixed finite $J_{K}$.
The lines and the shaded area have the same meaning as in
Fig.~\ref{fig:phase-K}.}
\label{fig:phase-SP}
\end{figure}

These results specify the phase diagram. It is hard to show the diagram in the
full parameter space $(J_{K}$, $J_{SP}$, $K_{\tau})$, as it would require a
three-dimensional graph. Instead, we plot two typical cross-sections,
corresponding to the cut at a finite constant value of $J_{SP}$
(Fig.~\ref{fig:phase-K}) and at a finite constant value of $J_{K}$
(Fig.~\ref{fig:phase-SP}).

One can also easily generalize these phase diagrams
to the case when $J_{SP}<0$ (see Appendix~\ref{sec:appen-Ising}).
There is a complete symmetry with respect to the change of
the sign of $J_{SP}$, except that one will have to replace
spin-Peierls phase with Ising one on the phase diagrams.

\section{Conclusions}

We have studied a one-dimensional Kondo lattice model that consists of a
one-dimensional electron gas coupled to a spin-$1/2$ chain through Kondo
interaction. The phase diagram is quite rich and contains phases with
either a large Fermi momentum or a small Fermi momentum.

When both nearest-neighbor and a significant next-nearest-neighbor
interactions are present in the isotropic spin chain and
Kondo interaction is small, the system is in the spin-Peierls phase, which is
characterized by broken translational symmetry due to the presence of spin
dimers and by small Fermi momentum $k_{F}$ of conduction electrons.

Increasing the strength of Kondo coupling constant, we have found 
indications that the system enters a coexistence phase,
containing both the dimers of Kondo singlets and the dimers of the 
spins in the spin chain. In this phase the translational symmetry is
also broken, but the conduction electrons are part of the system 
with large Fermi momentum $k_{F}^{\ast}$. Additionally, the spin 
gap encompasses conduction electrons as well. 

As we further increase Kondo interaction, the system enters
a pure Kondo-dimer phase with no translational symmetry broken and with
large Fermi momentum $k_{F}^{\ast}$. We were unable to determine the
precise nature of the transitions either between the spin-Peierls phase
and the coexistence phase or between the coexistence phase and the 
pure Kondo phase, due to the inherent limitations of the
renormalization-group approach.

We have also studied the model in which the spin chain contains an
$XY$ anisotropy. For sufficiently weak Kondo interaction, the system
is always in a Luttinger-liquid state, characterized by small Fermi
momentum $k_{F}$ and no breaking of translational invariance. Increasing
Kondo interaction triggers a \emph{continuous\/} Kosterlitz-Thouless
transition into the Kondo phase. We have calculated the dependence
of the spin gap on Kondo interaction near the transition and have found 
that it is exponentially small. We have also identified a correlation 
function that logarithmically diverges at zero frequency at
the transition.

The possibility of a continuous transition from small to large Fermi momenta
looks puzzling at first sight. In our case, the weight of the Kondo resonance
is characterized by the Kondo-dimer order parameter. The fact that the
transition is continuous reflects the continuous onset of this weight at the
transition. However, even infinitesimally small order parameter is still a
macroscopic quantity. The electron count includes the localized spins as soon
as the Kondo-dimer order parameter is developed, but does not include them as
long as we are in the Luttinger liquid phase. A jump in the Fermi momentum
then takes place at the transition point.

In dimensions higher than one, Fermi-momentum-changing transitions
have been discussed in the literature. One such transition
arises in an extended dynamical mean field treatment of the Kondo
lattice,\cite{Si01,Zhu03} between a large-Fermi-momentum 
paramagnetic metal phase and a small-Fermi-momentum 
antiferromagnetic metal phase. Here the transition is continuous and
is accompanied by a logarithmically divergent local spin susceptibility.
The transition is locally quantum critical, in the sense that
Kondo resonances are destroyed at the transition.
A related Fermi-momentum-changing transition has been discussed
in certain large-$N$ limit of frustrated Kondo
lattice systems,\cite{Senthil} between a large-Fermi-momentum
paramagnetic metal phase and a fractionalized small-Fermi-momentum
paramagnetic phase. The asymptotically exact study we have carried
out in the one-dimensional Kondo lattice model reveals a 
Fermi-momentum-changing transition that bears strong similarities
to the local quantum criticality proposed earlier for Kondo lattice
systems in dimensions higher than one.

We close by noting that some of the phases we have discussed for
the purely one-dimensional Kondo lattice may be of direct
experimental significance. For instance, the organic material
$\mathrm{(perylene)_2[Pt(S_2C_2(CN)_2)_2]}$ is believed to be 
a realization of a quasi-one-dimensional Kondo lattice;\cite{Bourbonnais}
the system also displays spin-Peierls ordering.
It would be very interesting to study the Fermi-surface properties 
of this material, as well as the phase transitions of the system by
tuning, say, pressure.

\begin{acknowledgments}
We would like to thank K.~Damle, S.~Sachdev, A.~M. Tsvelik 
and O.~Zachar for useful discussions. This work has
been supported by Robert A.~Welch Foundation, Research Corporation, 
TcSAM, and the NSF Grant No. DMR-0090071.
\end{acknowledgments}

\appendix

\section{$XY$ and Ising anisotropy.}
\label{sec:appen-Ising}

In our study of the Lagrangian Eq.~(\ref{eqn:L_anisotropic}), we were assuming
that the spin chain had $XY$ anisotropy. In this appendix we will generalize
our results for Ising anisotropy as well.

\begin{figure}[tbh!]
\begin{center}
\includegraphics[width=3.25in]{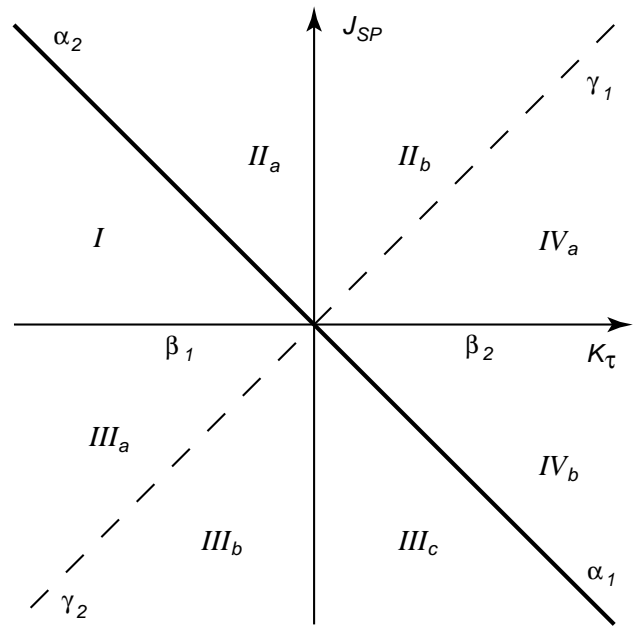}
\end{center}
\caption{$XY$ and Ising anisotropy (cross-section at $J_{K}=0$).
The line with $SU(2)$ symmetry is $\alpha$,
the regions $II$ and $IV$ have $XY$ anisotropy,
and the regions $I$ and $III$ have Ising anisotropy.
For $J_{K} \rightarrow \infty$, the line with
$SU(2)$ symmetry becomes $J_{SP}$ axis.}
\label{fig:symmetry}
\end{figure}

First of all, let us determine the boundary between the regions with $XY$ and
Ising anisotropy, i.e., the area with $SU(2)$ symmetry. In the full phase
diagram in the space of $J_{SP}$, $J_{K}$, and $K_{\tau}$ this area would be a
surface. However, if we look at the cross-section for fixed $J_{K}$, this area
reduces to a line (Fig.~\ref{fig:symmetry}).

For $J_{K}=0$, this line ($\alpha$) actually consists of two pieces. One
($\alpha_{1}$) that separates regions $III_{c}$ and $IV_{b}$ has a stable
fixed point at the origin, corresponding to the isotropic Luttinger liquid.
One ($\alpha_{2}$) that separates regions $I$ and $II_{a}$ has a stable fixed
point at infinity, corresponding to spin-Peierls phase. The regions $I$ and
$II$ (both with $J_{SP}>0$) flow towards spin-Peierls fixed point, the region
$III$ flows towards Ising fixed point on the line $\gamma_{2}$ at infinity,
which corresponds to N{\'e}el order with broken Ising symmetry,
and the region $IV$ flows towards the line of Luttinger-liquid fixed points
$\beta_{2}$. This is a standard Kosterlitz-Thouless
picture.\cite{Jose77,Kogut79}

As $J_{K}\rightarrow\infty$, the line with $SU(2)$ symmetry rotates so that it
overlaps with $J_{SP}$ axis. Now in all regions the flow is towards the line
of Kondo-phase fixed points along $J_{SP}$ axis. For finite values of $J_{K}$,
spin-Peierls and Kondo fixed points still possess $SU(2)$ symmetry, while
Luttinger-liquid and Ising fixed points break it.

The continuous transition from the Luttinger-liquid to 
the Kondo phase is always in
the $XY$ region and it exists for any sign of $J_{SP}$, but only 
for $K_{\tau}>1$, which is why most of the calculations in this paper have been
performed assuming $XY$ anisotropy. The line segment $\gamma_{1}$ that was
separating Luttinger liquid from spin-Peierls phase for $J_{K}=0$ now splits
into two boundaries that surround the Kondo phase (Fig.~\ref{fig:phase-SP}).

As for the transition from Kondo to spin-Peierls phase, its study is more
complicated due to the mentioned possibility of the 
coexistence region. However, it is clear that 
the finite $J_K$ makes this transition 
(or the coexistence region) extend towards 
a finite range within region $I$ as well.
We have determined the location of the coexistence region
for Ising anisotropy with $J_{SP}>0$ and $K_{\tau}<1$, which is reflected in
Eq.~(\ref{eqn:dashed-line}).

What will happen for the remaining region with Ising anisotropy ($III$)? Let
us observe that the RG equations Eq.~(\ref{eqn:RG}) are invariant with respect
to the change of the sign of $J_{SP}$. This means that both of our phase
diagrams Figs.~\ref{fig:phase-K},\ref{fig:phase-SP} should remain the same, except
that now $J_{SP}$ will need to be replaced with $-J_{SP}$ and Ising order will
take place of spin-Peierls one. In the coexistence
region (now for Kondo and Ising phases) the electrons not only will become
spin-gapped, but also will break $SU(2)$ invariance so that Ising order will
effectively expand over the entire system.

\section{Small Fermi momentum in the decoupled Luttinger liquid}
\label{sec:appen-small-kF}

In this appendix, we consider the decoupled
Luttinger liquid phase in which the transverse component of the
Kondo coupling ($J_K^{f\perp}$) is irrelevant. We wish to establish
that the Fermi momentum is $k_F$. This statement is obviously
true if the longitudinal component of the Kondo coupling
($J_K^{fz}$) is absent, in which case the conduction electrons
would be completely decoupled from the spin chains when we reach
the fixed points. Below, we show that it remains valid even for
a finite $J_K^{fz}$-coupling.

The $J_K^{fz}$ term describes the forward-scattering interaction
between the $z-$components of the spins of the conduction electrons
and those of the spin-chain [cf. Eq.~(\ref{eqn:L-terms-c})].
It is marginal in the RG sense when $y^{\perp}=0$
[cf. Eq.~(\ref{eqn:RG-y_z})]. In addition, both $K_{\tau}$
and $K_s$ are also marginal since,
in the decoupled Luttinger liquid phase,
$y_{SP}=0$ as well.
We can then introduce the following diagonalization (mentioned already
in Section~\ref{sec:RG}):
\begin{subequations}
\label{eqn:diagonalization}
\begin{align}
\phi_{s}^{\prime}  & =a_{s}\phi_{s}+b_{s}\phi_{\tau},
\label{eqn:diag-a}\\
\phi_{\tau}^{\prime}  & =a_{\tau}\phi_{s}+b_{\tau}\phi_{\tau},
\label{eqn:diag-b}\\
v_{s}^{\prime}  & =c_{s}v_{s}+d_{s}v_{\tau},
\label{eqn:diag-c}\\
v_{\tau}^{\prime}  & =c_{\tau}v_{s}+d_{\tau}v_{\tau}.
\label{eqn:diag-d}
\end{align}
\end{subequations}
The transformation coefficients can be straightforwardly
derived for arbitrary values of $J_K^{fz}$. For simplicity,
however, we will write down the expressions that are valid 
only up to the second order in $J_K^{fz}$:
\begin{subequations}
\begin{align}
a_{s}  & =1-\frac{v_{\tau}}{4v_{s}}\frac{v_{s}^{2}+v_{\tau}^{2}}
{\left(  v_{s}^{2}-v_{\tau}^{2}\right)  ^{2}}K_{s}K_{\tau}
\left(  \frac{J_{z}}{2\pi }\right)  ^{2},\\
b_{s}  & =\frac{v_{s}K_{s}}{v_{s}^{2}-v_{\tau}^{2}}
\left(  \frac{J_{z}}{2\pi }\right)  ,\\
c_{s}  & =1+\frac{v_{s}v_{\tau}}
{\left(  v_{s}^{2}-v_{\tau}^{2}\right)  ^{2} }
K_{s}K_{\tau}\left(  \frac{J_{z}}{2\pi}\right)  ^{2},\\
d_{s}  & =-\frac{v_{s}^{2}+v_{\tau}^{2}}{2\left(  v_{s}^{2}
-v_{\tau}^{2}\right)  ^{2}}K_{s}K_{\tau}
\left(  \frac{J_{z}}{2\pi}\right)  ^{2} .
\end{align}
\end{subequations}
The expressions for $a_{\tau}$, $b_{\tau}$,
$c_{\tau}$, and $d_{\tau}$ have the same forms, except that
the subscripts $s$ and $\tau$ are exchanged.

The single electron Green's function,
$G_{\sigma} (x,t) \equiv -\left\langle T_{t}c_{\sigma}(x,t)
c_{\sigma}^{\dag}(0,0)\right\rangle$, can now be calculated
using the bosonization form: 
$c_{\sigma} (x) = \sum_{r = \pm } c_{r\sigma}(x)$,
where $r = \pm $ labels the right/left moving parts
and 
\begin{equation}
c_{r \sigma}(x) = \frac{1}{\sqrt{2 \pi a}}
e^{i r k_{F} x} 
F_{r\sigma} 
\exp\left\{  -i r \Phi_{r\sigma} (x) \right\} .
\end{equation}
Here, $\Phi_{r\sigma} (x) \equiv \sqrt{\pi}
\left[  \phi_{\sigma} + r \theta_{\sigma} \right]$.
For definiteness, we consider the 
Green's function of a right-moving electron,
$G_{+\sigma}(x,t) \equiv -\left\langle T_{t}c_{+\sigma}(x,t) 
c_{+\sigma}^{\dag} (0,0)\right\rangle $.
Using $\Phi_{+\sigma} = (1/\sqrt{2})\Phi_{c,+}
+ (\sigma / \sqrt{2}) \Phi_{s,+} $
and the properties of the Klein 
factors described in the main text, 
we can see that 
$G_{+}(x,t)$ factorizes into a charge part and a spin part,
\begin{align}
G_{+\sigma}(x,t) &= 
-\frac{1}{2 \pi a}
e^{i k_{ F} x} G_{c,+}(x,t) G_{s,+\sigma}(x,t) , \nonumber\\
G_{c,+}(x,t) &= \left\langle T_{t} \exp\left\{  -\frac{i}{\sqrt{2}}
\Phi_{c,+} (x,t) \right\}\right.  \nonumber\\
& \qquad\times\exp\left.\left\{ \frac{i}{\sqrt{2}} \Phi_{c,+} (0,0)
\right\} \right\rangle , \\
G_{s,+\sigma}(x,t) &= \left\langle T_{t} \exp\left\{ -\frac{i\sigma }
{\sqrt{2}} \Phi_{s,+} (x,t) \right\} \right. \nonumber \\
& \qquad\times\exp\left.\left\{ \frac{i\sigma }{\sqrt{2}} \Phi_{s,+} (0,0)
\right\} \right\rangle . \nonumber
\end{align}
The charge part,
$G_{c,+}(x,t)$, is determined by the 1DEG alone.
Because of the $J_K^{fz}$ coupling,
the spin part, $G_{s,+\sigma}(x,t)$, \emph{does\/} involve
the spins of both the 1DEG and the spin-chain.
This $J_K^{fz}$ coupling, however, can be handled by
going to the diagonalized basis
$\phi_{s}^{\prime}$ and 
$\phi_{\tau}^{\prime}$ introduced in 
Eq.~(\ref{eqn:diagonalization}).
Since the diagonalization affects only the $q \sim 0$ 
spin component of the spin chain, going to the 
primed basis does not introduce any oscillatory factor
in the spatial dependence of 
$G_{s,+\sigma}(x,t)$. As a result, the single-electron
Green's function $G_{+\sigma}(x,t)$ is a product 
of $e^{ik_F x}$ and two factors which
decay algebraically in $x - v_c t$ and $x - v_s t$,
respectively. The Fermi momentum, hence, is $k_F$.

The key to the above reasoning is that only the forward 
scattering component of the longitudinal 
Kondo interaction is marginal. 
All other components, including those that
involve either the $q=\pm \pi$ mode of the spin
chain or the $q = \pm 2k_{\rm F}$ spin mode of the 
1DEG [cf. Eqs.~(\ref{eqn:s-decompose},\ref{eqn:tau-decompose})]
are irrelevant. In particular, the irrelevant nature
of the interactions involving the staggered
moment of the spin chain leaves no room for the single
electron Green's function to contain any
Fermi momentum other than $k_{\rm F}$.

\end{document}